%% file: kuberski_lat19.tex
\documentclass{PoS}
\pdfoutput=1

\usepackage[utf8]{inputenc}

\usepackage{xcolor}

\usepackage{mathtools}

\usepackage{booktabs}

\usepackage{wrapfig}

\usepackage{url}

\newcommand{\mr}[1]{\mathrm{#1}}

\graphicspath{{plots/}}

\title{Towards the determination of the charm quark mass on $N_\mathrm{f}=2+1$ CLS ensembles}

\ShortTitle{Towards the determination of the charm quark mass on $N_\mathrm{f}=2+1$ CLS ensembles}

\author{
	\begin{minipage}[b]{0.4\linewidth}
		\includegraphics[height=2.5\baselineskip]{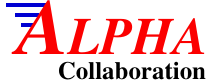}
	\end{minipage}
	\hfill \\
	\hfill\parbox{37.5mm}{\vspace{-2.25cm}\raggedleft\footnotesize\it%
		MS-TP-19-22
	}
}

\author{Jochen Heitger, Fabian Joswig and \speaker{Simon Kuberski} 	
	\\
	Westfälische Wilhelms-Universität Münster, Institut für Theoretische Physik, \\ Wilhelm-Klemm-Straße 9, 48149 Münster, Germany \\
	E-mail: \email{simon.kuberski@uni-muenster.de}}

\abstract{We present the current status of our lattice QCD determination of the charm quark mass using $N_\mathrm{f}=2+1$ dynamical, non-perturbatively $\mathrm{O}(a)$ improved Wilson fermions. A subset of CLS ensembles with five different lattice spacings along the $\mr{Tr}[M_\mr{q}]=\text{const.}$ trajectory is used. For the computation of the correlation functions involving valence charm quark propagators, we employ distance preconditioning to gain the necessary precision. To stabilize the extrapolations to the physical point, we consider different definitions of the bare charm quark mass and corresponding renormalization procedures.}

\FullConference{37th International Symposium on Lattice Field Theory - Lattice2019\\
	16-22 June 2019\\
	Wuhan, China}

\begin{document}
	
	\section{Introduction}\vspace{-.5em}
	As fundamental parameters of the Standard Model, the masses of the
	quarks are of great phenomenological interest. 
	Particularly the precise knowledge of the charm and bottom quark mass
	values is crucial for the search for new physics, because decay rates
	and branching ratios of the Higgs boson depend critically on the masses
	of these heavy quarks (see, e.g., \cite{Petrov:2015jea}) and at future 
	lepton colliders it will be possible to measure their Yukawa couplings 
	to very high accuracy. 	
	
	As a consequence of confinement, any determination of quark masses must
	relate them to the observable, low-energy hadronic world and thus requires
	a reliable quantitative control over this genuinely non-perturbative
	regime of QCD.
	Therefore, lattice QCD has emerged as an ideal calculational tool to
	provide precise quark mass results.
	Nevertheless, some difficulties have to be overcome.
	Apart from properly dealing with the inherent renormalization scheme and
	scale dependence of quark masses, mass dependent cut-off effects can
	become sizable towards the charm sector, when Wilson fermions are
	considered, such that full $\mathrm{O}(a)$ improvement and the use of 
	small lattice spacings are needed.
	
	Here we report the status of our ongoing computation to determine the 
	mass of the charm 	quark in $N_\mathrm{f}=2+1$ lattice QCD with Wilson 
	fermions, which applies
	recent non-perturbative results for the (scale dependent) quark mass
	renormalization factor \cite{Campos:2018ahf}, as well as for some of the
	improvement coefficients (multiplying the quark mass dependent terms
	involved) \cite{deDivitiis:2019xla}.
	A calculation of the light and strange quark masses for the same lattice
	discretization is also under way \cite{Bruno:2019xed}. 
	
	\section{Setup}\vspace{-.5em}
	\begin{wrapfigure}[12]{R}{0.45\textwidth}
		\vspace{-2.5em}
		\centering
		\includegraphics[width=\linewidth]{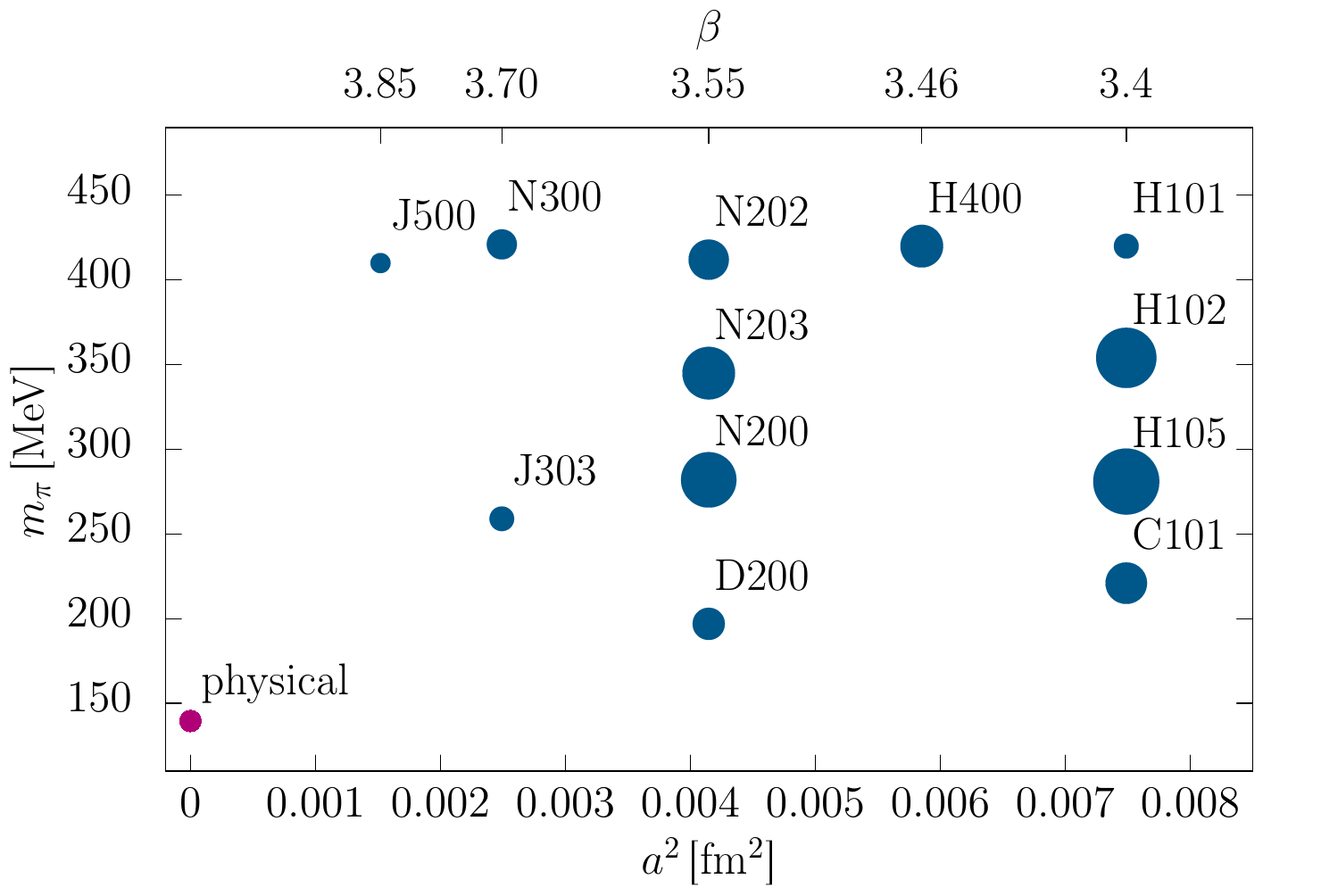}
		\caption{Status of the measurements. The area of the circles is proportional to the number of measurements. Ensemble ids are given in \cite{Bruno:2014jqa, Mohler:2017wnb}.}
		\label{fig:ensembles}
	\end{wrapfigure}

	We work on CLS ensembles with ${N_\mr{f}=2+1}$ flavors of $\mathrm{O}(a)$
	improved Wilson fermions and Lüscher-Weisz gluons \cite{Bruno:2014jqa, Mohler:2017wnb}. 
	In this work we constrain ourselves to the mass trajectory with $\mr{Tr}[M_\mr{q}]=2m_\mr{l}+m_\mr{s}=\text{const.}$ All considered 
	ensembles feature open boundary conditions in time to allow for a proper 
	sampling of the topological charge. Measurements for five different lattice 
	spacings $a$ down to $\approx 0.04\,\mr{fm}$ and pion masses  $m_\pi$ down 
	to $\approx 200\,\mr{MeV}$ are included. An overview of the considered ensembles 
	is given in Fig.~\ref{fig:ensembles}. In the near future, we will increase the 
	statistics on the finest ensembles and add ensembles with smaller pion masses 
	for some couplings.
	
	We work in a partially quenched setup, where the charm quark only enters 
	in the valence sector.
	To determine effective meson masses and the quark masses, we calculate 
	the two-point correlation functions
	\begin{align}
	f_O^{rs}(x_0,y_0)=-\frac{a^6}{L^3}\sum_{\vec{x},\vec{y}}\langle O^{rs}(x_0,\vec{x})P^{rs}(y_0,\vec{y})\rangle\,, \quad O^{rs}=\bar{\psi}^r(x)\,\Gamma\,\psi^s(x)\,,
	\end{align}
	where $P$ is the pseudoscalar density, for all possible flavor combinations 
	$rs$ and various Dirac structures $\Gamma$. The sources are placed at the 
	boundaries, i.e., at $y_0=a$ and $y_0=T-a$. To decrease the statistical 
	error, we use 16 $U(1)$ noise sources per time slice.
	
	Bare quark masses can be calculated from the $\mathrm{O}(a)$ improved PCAC 
	relation via
	\begin{align}
	am_{rs}(x_0)=\frac{\tilde{\partial}_0^{\phantom{*}} f_{\mathrm{A}_0}^{rs}(x_0)+ac_\mathrm{A}\partial^*_0\partial_0^{\phantom{*}}f_\mathrm{P}^{rs}(x_0)}{2f_\mathrm{P}^{rs}(x_0)}\,, \label{eq:PCAC}
	\end{align}
	where $A_0$ is the temporal component of the axial current and 
	$\tilde{\partial}_0^{\phantom{*}}$, $\partial^*_0$ and $\partial_0^{\phantom{*}}$
	are lattice representations of the central, backward and forward
	derivative. The improvement coefficient $c_\mr{A}$ is known non-perturbatively
	from ref.~\cite{Bulava:2015bxa}.
	
	We consider two heavy valence quarks with masses above and below the physical 
	charm quark mass. For both choices we determine the effective masses of the 
	pseudoscalar mesons corresponding to $D$ and $D_\mr{s}$ and of the vector 
	mesons corresponding to $D^*$ and $D_\mr{s}^*$. The hopping parameter 
	$\kappa_\mathrm{c}$ for a physical charm quark is determined on each ensemble 
	by an interpolation of the meson masses to their physical values. In principle, 
	each of the aforementioned mesons could be chosen to fix $\kappa_\mathrm{c}$ 
	and the resulting values from different choices are expected to differ only 
	by cut-off effects. 
	
	Since $\mr{Tr}[M_\mr{q}]$ is kept constant for all our ensembles, we can choose 
	the flavor-averaged meson mass $M=\frac{1}{3}({2m_\mathrm{D}}+{m_\mathrm{D_s}})$ 
	for the calibration and expect the dependence on the light quark masses to 
	be rather mild. In addition, we can use insights from heavy quark effective 
	theory \cite{Falk:1992wt, Neubert:1993mb} concerning the heavy meson masses, 
	to remove short-distance effects from spin-interactions of the heavy quark. 
	Therefore, the appropriate spin average on top of the flavor average, leading 
	to the average mass
	\begin{align}
	M=\frac{1}{12}({6m_{\mathrm{D}^*}}+{2m_\mathrm{D}}+{3m_{\mathrm{D}_\mathrm{s}^*}}+{m_\mathrm{D_s})}\,,
	\end{align}
	is expected to decrease cut-off effects in the tuning procedure. At the same 
	time, however, the statistical error on the vector meson masses is significantly 
	larger than the error on the pseudoscalar meson masses. This could propagate into 
	the final result. We thus consider both possibilities and judge the quality of 
	the chiral-continuum extrapolations afterwards.
	
\begin{figure}[!htb]
		\centering
		\begin{minipage}{.49\textwidth}
			\centering		
			\includegraphics[width=\linewidth]{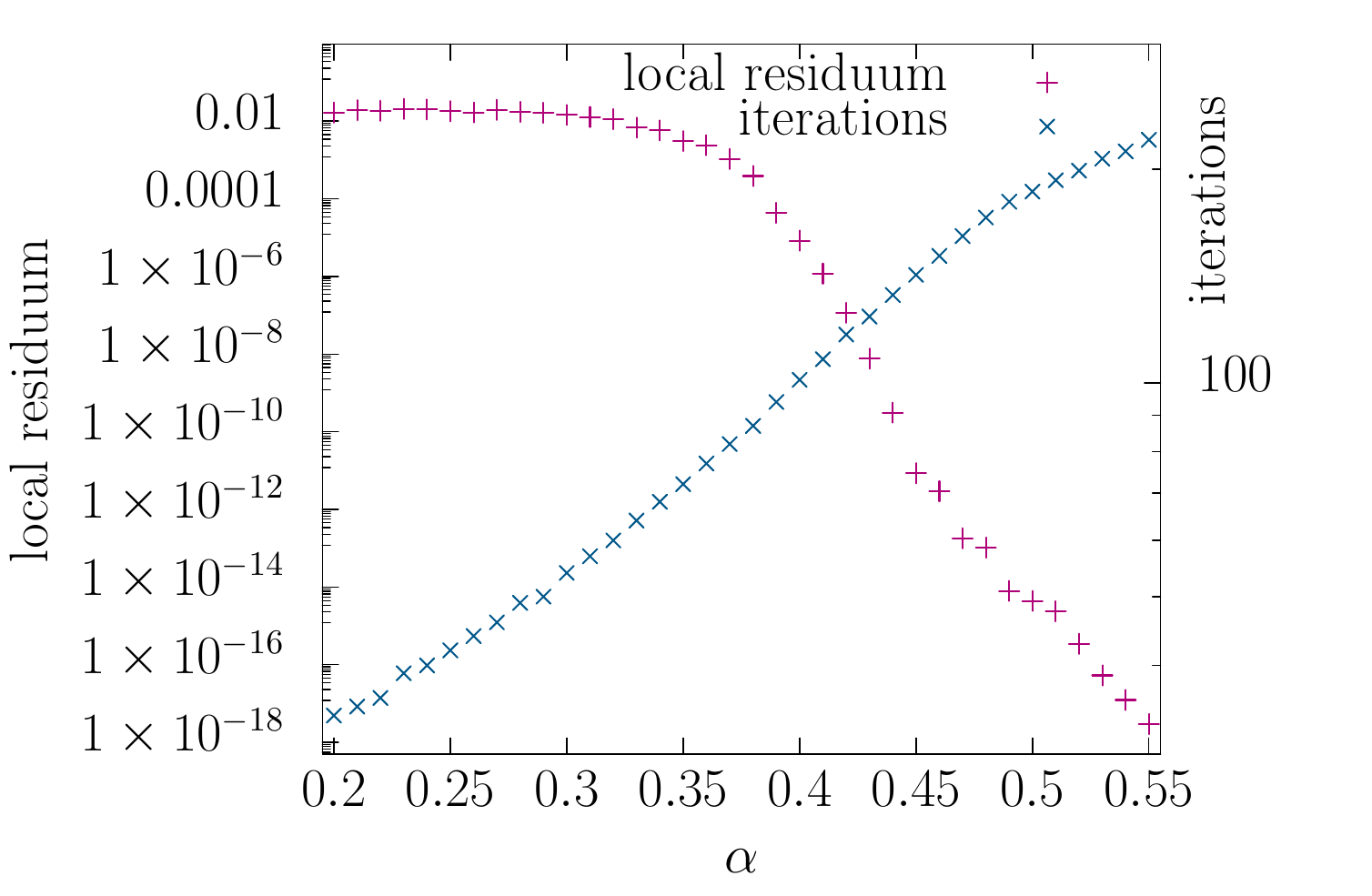}
		\end{minipage}
		\begin{minipage}{0.49\textwidth}
			\centering
			\includegraphics[width=\linewidth]{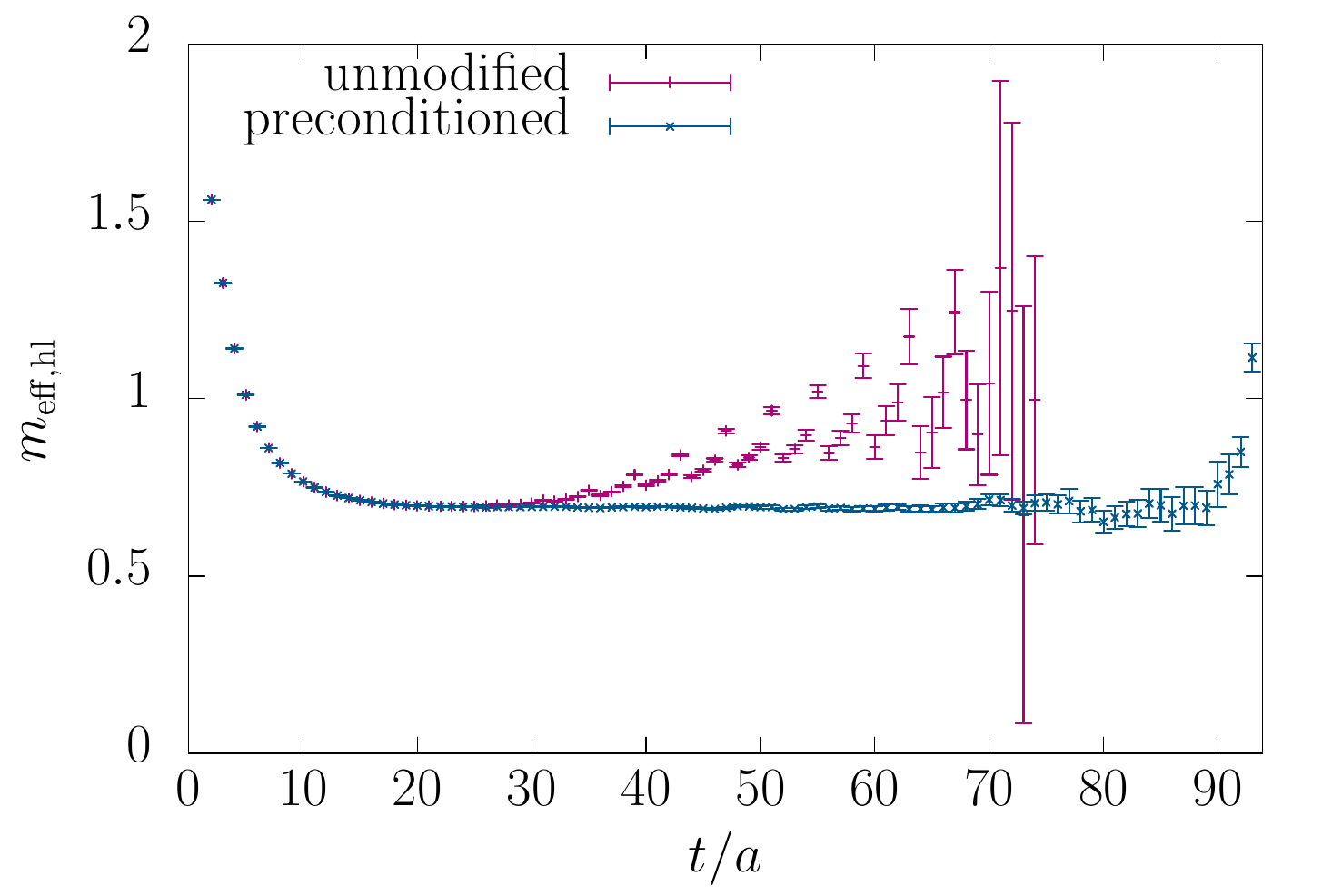}
		\end{minipage}
		\caption{\textit{Left:} The local residuum at $x_0=\frac{7}{8}T$ and the number of iterations to reach the global residuum depending on the preconditioning parameter $\alpha$ on one configuration of the H400 ensemble. \textit{Right:} Effective mass of the $D$ meson determined on all configurations with and without distance preconditioning. The rise in $m_\mathrm{eff, hl}$ at late time slices for the preconditioned mass is due to the boundary at $T/a=95$.
			\label{fig:dp}}
		\vspace{-1em}
	\end{figure}

	On top of the expected cut-off effects, heavy quarks can also introduce numerical 
	difficulties. When the iterative solution of the Dirac equation is based on a 
	global residuum 
	\begin{align}
	\left|\sum_z \mathbb{D}_{x,z} \:S_\mathrm{h}(z)-\eta(x)\right| < r_\mathrm{gl}\quad \text{with}\quad 
	\mathbb{D} = D[U]+m_\mathrm{h}\,,
	\end{align}
	and the mass $m_\mr{h}$ is heavy, time slices 
	far away from the source are exponentially suppressed, leading to incorrect 
	solutions at late times \cite{Juttner:2005ks}. Distance preconditioning 
	\cite{deDivitiis:2010ya} can be used to achieve numerically accurate results 
	at all time slices. Instead of the original Dirac equation, the preconditioned 
	system \cite{Collins:2017iud}, in matrix notation
	\begin{align}
	(P\,\mathbb{D}\,P^{-1})(PS)=(P\eta),\quad  P=\mathrm{diag}(p_i), \quad p_i=\exp\left(\alpha\, |y_0-x_0^{(i)}|\right)\,,
	\end{align}
	is solved and the desired solution is obtained by the multiplication of the 
	preconditioned solution with the inverse of the preconditioning matrix $P$. 
	With an appropriate choice for the parameter $\alpha$, the exponential decay 
	of the propagator is counteracted such that, effectively the heavy quark acts 
	as a light quark. To keep the additional cost under control, $\alpha$ has to be 
	tuned carefully. Figure \ref{fig:dp} shows the local residuum 
	\begin{align}
	r_\mathrm{loc}(x_0,y_0) \equiv \frac{\left|(\mathbb{D}\,S_\mathrm{h})(x_0, y_0)-\eta(x_0,y_0)\right|}{\left|S_\mathrm{h}(x_0,y_0)\right|}\,,\quad \text{with}\quad y_0=a,\enspace x_0=\frac{7}{8}T\,,
	\end{align}
	against $\alpha$ together with the number of iterations to reach 
	$ r_\mathrm{gl} = 10^{-8}$. When $\alpha$ is increased above some threshold, 
	$r_\mathrm{loc}$ starts to decrease exponentially, while the cost increases 
	exponentially.
	
	The effect of the preconditioned solver can be seen on the right hand side 
	of Fig.~\ref{fig:dp}, where we show the effective mass of the pseudoscalar 
	heavy-light meson on the H400 ensemble for the preconditioned and the standard 
	solver. Without preconditioning, the identification of a plateau is ambiguous 
	since the effect of the numerical instabilities already dominates at comparably 
	small times, leading to possibly large systematic uncertainties. We use the 
	implementation of the distance preconditioned SAP-GCR solver \cite{Collins:2017iud} 
	in the open source package \texttt{mesons} \cite{mesons}.
	
	\section{Renormalized quark masses}\vspace{-.5em}
	With $\kappa_\mathrm{c}$ at hand, we can interpolate the PCAC masses involving 
	a heavy propagator to obtain the bare charm quark mass. To determine physical 
	and $\mathrm{O}(a)$ improved masses, we need to renormalize and improve the bare 
	quark masses. Taking the relevant formulae from refs.~\cite{Campos:2018ahf, deDivitiis:2019xla}, 
	we arrive at
	\begin{align}
	M^\mathrm{RGI}_{rs} &= \frac{M}{\overline{m}(\mu_\mathrm{had})}m_{rs,\mathrm{R}}\equiv
	\frac{M}{\overline{m}(\mu_\mathrm{had})} \frac{Z_\mathrm{A}}{Z_\mathrm{P}(\mu_\mathrm{had})} m_{rs} \left[1+\frac{({b_\mathrm{A}-b_\mathrm{P}})}{{Z}}am_{rs}-b_M a\mathrm{Tr}[M_\mathrm{q}] \right]
	\end{align}
	as general formula for a non-degenerate renormalized renormalization group 
	invariant (RGI) quark mass. The running factor ${M}/\overline{m}$ to evolve 
	the mass from the hadronic scale $\mu_\mathrm{had}$ to the RGI value, as well 
	as the non-perturbatively determined renormalization constant for the pseudoscalar 
	density $Z_\mathrm{P}$, are available from \cite{Campos:2018ahf}. The renormalization 
	constant for the axial current $Z_\mr{A}$ was determined in \cite{Bulava:2016ktf, DallaBrida:2018tpn}. We work with the value from ref.~\cite{DallaBrida:2018tpn} 
	because of its smaller statistical uncertainties. The combination of the improvement 
	coefficients $(b_\mathrm{A}-b_\mathrm{P})$ and the normalization constant ${Z={Z_\mathrm{m}Z_\mathrm{P}}/{Z_\mathrm{A}}}$ have recently been determined 
	non-perturbatively \cite{deDivitiis:2019xla}. They allow for correcting for the 
	valence quark dependent piece of the $\mathrm{O}(a)$ effects, which are expected 
	to be dominating for valence quark in the charm region. Whereas non-perturbative 
	results for $r_\mathrm{m}$ are available in \cite{Bali:2016umi}, the full factor 
	multiplying the sum of the sea quark masses, defined as
	\begin{align}
	b_M \equiv ({r_\mathrm{m}-1})\frac{({b_\mathrm{A}-b_\mathrm{P}})}{N_\mathrm{f}}+({\overline{b}_\mathrm{A}-\overline{b}_\mathrm{P}})\,,
	\end{align}
	is not known non-perturbatively so far. Therefore we neglect this subleading piece 
	of the improvement and investigate the possibility of residual $\mathrm{O}(a)$ effects 
	in the continuum extrapolation.
	
	By choosing different combinations of flavors $rs$, we can arrive at various definitions 
	of the renormalized charm quark mass. Imposing two mass degenerate flavors c and 
	$\mathrm{c}'$ at the mass of the physical charm quark allows us to employ the clean 
	signal of the PCAC mass from the heavy-heavy propagator to calculate the RGI mass via
	\begin{align}
	M^\mathrm{RGI}_\mathrm{c} &= \frac{M}{\overline{m}(\mu_\mathrm{had})}m_{\mathrm{cc',R}}\,. \label{eq:mrgi_cc}
	\end{align}
	Since we expect the mass dependent cut-off effects to be rather large for this choice, 
	we also consider the definition based on the light-heavy correlation functions,
	\begin{align}
	2m_{\mathrm{lc,R}}-m_{\mathrm{ll',R}} \equiv 2\,\frac{m_{\mathrm{c,R}}+m_{\mathrm{l,R}}}{2}-\frac{m_{\mathrm{l,R}}+m_{\mathrm{l,R}}}{2} = m_\mathrm{c,R}\,,
	\end{align}
	and the analogous expression from the strange-heavy correlation functions, where the 
	non-degenerate quark masses have been defined in eq.~(\ref{eq:PCAC}). Combining both 
	to a flavor averaged mass, to reduce the slope in the chiral extrapolation, we arrive at 
	\begin{align}
	M^\mathrm{RGI}_\mathrm{c} &= \frac{M}{\overline{m}(\mu_\mathrm{had})}\frac{1}{3}\left[2(2m_{\mathrm{lc,R}}-m_{\mathrm{ll',R}})+(2m_{\mathrm{sc,R}}-m_{\mathrm{ss',R}})\right] \label{eq:mrgi_ls}
	\end{align}
	as second definition for a renormalized charm quark. Both definitions can be used to 
	determine a chiral-continuum extrapolated quark mass. This leads to a reduction of 
	systematic effects.

	\section{Preliminary results}
\begin{figure}[!htb]
		\centering
		\begin{minipage}{.49\textwidth}
			\centering		
			\includegraphics[width=\linewidth]{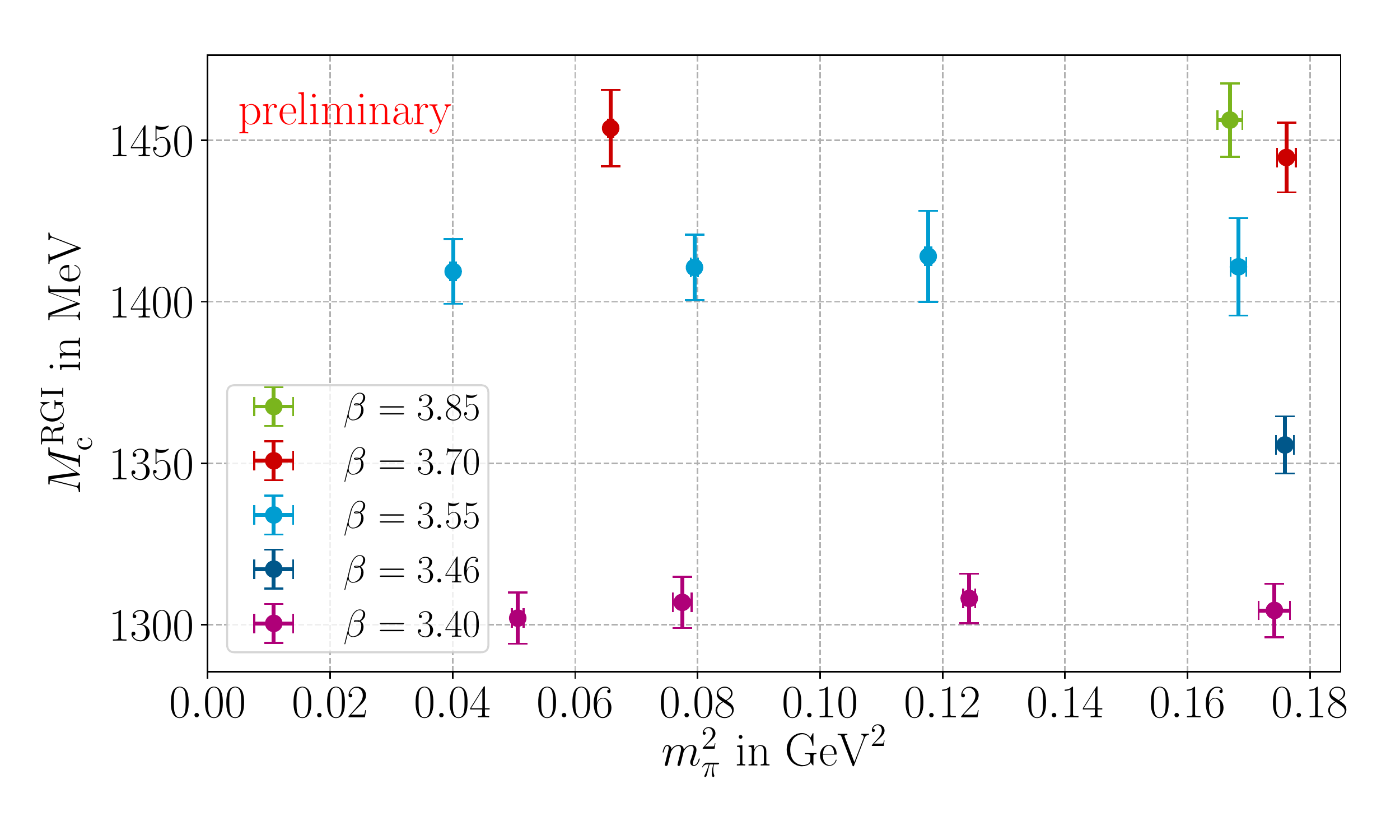}
		\end{minipage}
		\begin{minipage}{0.49\textwidth}
			\centering
			\includegraphics[width=\linewidth]{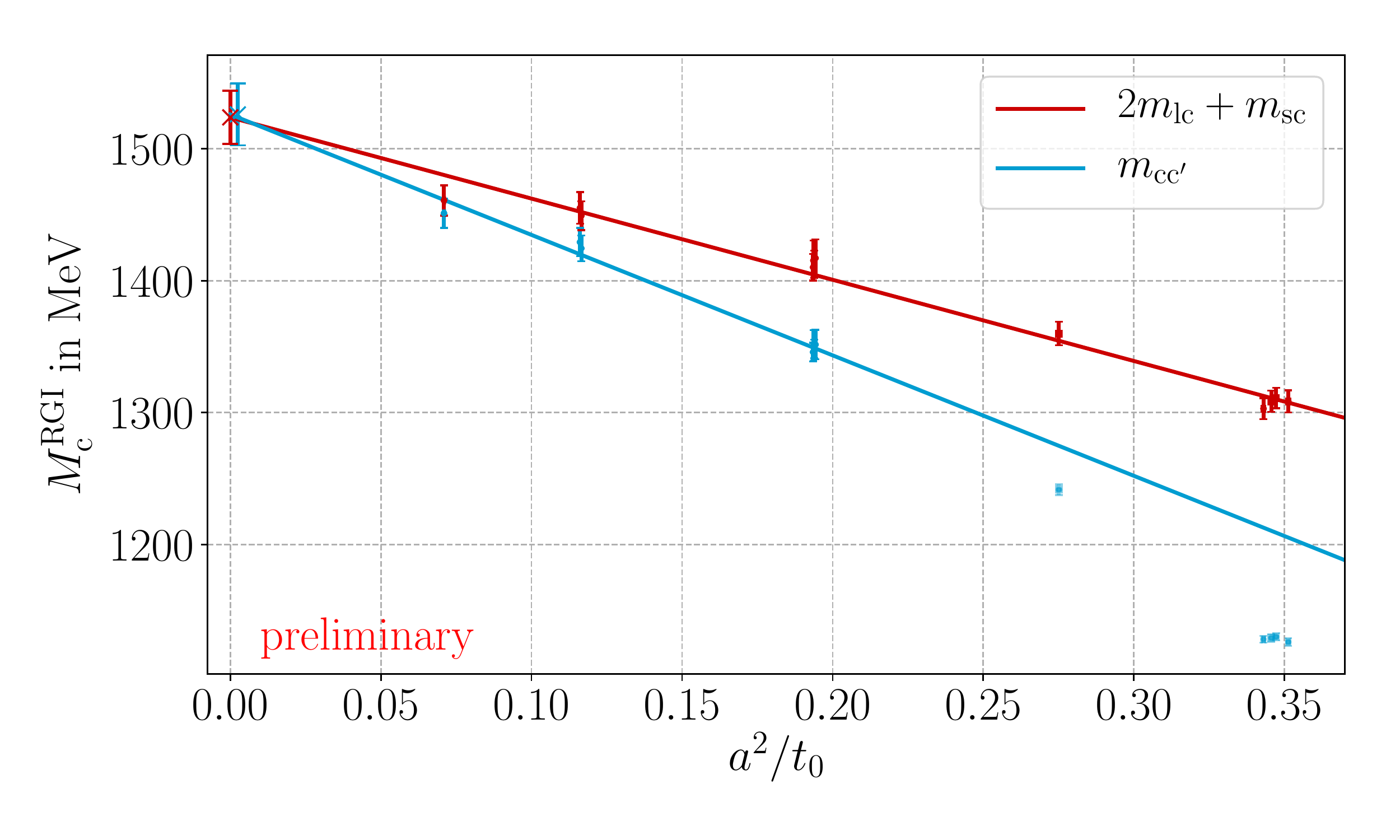}
		\end{minipage}
		\caption{\textit{Left:} Overview of the RGI charm quark masses evaluated on the considered ensembles depending on the pion mass. Different bare couplings and thus lattice spacings are indicated with different colors. \textit{Right:} Chiral-continuum fits for both definitions of the charm quark mass. The higher-dimensional fits and the data points are projected onto the plane of the chiral trajectory.} 
		\label{fig:chiral-continuum}
		\vspace{-.8em}
	\end{figure}

\vspace{-.5em}
	In Figure \ref{fig:chiral-continuum} we present the preliminary results of our 
	analysis. On the left hand side the RGI charm quark mass determined from the 
	definition in eq.~(\ref{eq:mrgi_ls}) against the pion mass $m_\pi$ is shown for 
	the five values of the bare inverse coupling $\beta$. From the ensembles at 
	$\beta=3.40$ and $\beta=3.55$ it can be seen, that there is no significant 
	dependence of $M^\mathrm{RGI}_\mathrm{c}$ on the light quark masses. At the 
	same time, the cut-off effects are rather large.
	
	On the right hand side of Fig.~\ref{fig:chiral-continuum} the preliminary 
	chiral-continuum fits for both definitions (\ref{eq:mrgi_cc}) and (\ref{eq:mrgi_ls}) 
	are shown. As expected, the masses based on the heavy-heavy current seem to 
	suffer from larger cut-off effects. At this stage of the analysis we perform 
	a fit to the polynomial form
	\begin{align}
	M^\mathrm{RGI}_\mathrm{c}\left(t_0\delta_M^2, \frac{a^2}{t_0}\right)=c_0\left(1+c_1t_0\delta_M^2\right)\left(1+c_2\frac{a^2}{t_0}\right) \quad \text{with} \quad t_0\delta_M^2 = t_0(m_K^2-m_\pi^2)\,, \label{eq:fitformula}
	\end{align}
	with the fit parameters $c_i$ which parameterize the leading chiral and 
	cut-off effects. The chiral point is defined at the physical value of 
	$t_0\delta_M^2$. The gluonic quantity $t_0$ is defined from the gradient 
	flow and its physical value has been computed in ref.~\cite{Bruno:2016plf}. 
	No linear dependence on $a$ can be resolved. For the two coarsest lattice 
	spacings, we observe higher-order effects in $M^\mathrm{RGI}_\mathrm{c}$ 
	from definition (\ref{eq:mrgi_cc}) (blue points). Therefore we decide to 
	exclude these points from our fit with the ansatz~(\ref{eq:fitformula}). 
	For the definition~(\ref{eq:mrgi_ls}) (red points), all ensembles are taken 
	into account.
	
	As it can be seen from Fig.~\ref{fig:chiral-continuum}, both definitions 
	nicely coincide in the continuum limit. Although the different PCAC masses 
	are partly correlated, we see this as evidence that possible systematic errors 
	in the fit are under good control.
	
	\section{Outlook}\vspace{-.5em}
	We have presented preliminary results of our determination of the RGI 
	charm quark mass on the $N_\mr{f}=2+1$ CLS ensembles. To arrive at stable 
	plateaus for the heavy-light mesons, we used a distance preconditioned 
	solver. Our continuum extrapolations are monitored by extrapolating 
	several definitions of the renormalized charm quark mass. Barely any 
	effect of the sea quarks on the charm quark mass can be seen in our 
	parameter region down to $m_\pi=200\,$MeV. Although we refrain from 
	quoting a number for $M^\mathrm{RGI}_\mathrm{c}$ at this preliminary 
	stage, we note that our extrapolated value is close to the FLAG average 
	quoted in \cite{Aoki:2019cca}. With the current status, we can foresee 
	that the final precision will almost reach the $\approx1$\% limit dictated 
	by the uncertainty on the running factor.
	
	To arrive at final results, further steps will be done: As it can be 
	concluded from Fig.~\ref{fig:ensembles}, we have to increase the 
	statistics on the most demanding ensembles. Additional ensembles on 
	the $\mr{Tr}[M_\mr{q}]=\text{const.}$ trajectory are available for 
	three lattice spacings and will be considered. This effort is ongoing. 
	Especially the ensembles at the finest lattice spacing will help to 
	stabilize the fit to the continuum. Other definitions of the renormalized 
	quark mass, e.g., from the bare current quark mass and the ratio-difference 
	method \cite{Durr:2010aw} can be explored to investigate possible systematic 
	effects. The impact of a combined fit of several definitions will be studied. 
	Although we do not expect any finite-volume effects on our charm observables, 
	we will explicitly check this for one representative point in the parameter space.
	
	As it is described in ref.~\cite{Bruno:2016plf}, our ensembles deviate 
	slightly from the chiral trajectory chosen for the 
	$\mr{Tr}[M_\mr{q}]=\text{const.}$ trajectory. This can be corrected by 
	a slight shift in the sea quark masses. In order to incorporate the effect 
	on our observables, we calculated the derivatives of the correlation 
	functions with respect to a shift in the sea quark masses.
	
	\section*{Acknowledgments}
	\vspace{-.8em}
	We would like to thank Mattia Bruno, Sara Collins, Kevin Eckert, 
	Tomasz Korzec and Anastassios Vladikas for helpful discussions.
	This work is supported by the Deutsche Forschungsgemeinschaft (DFG) 
	through the Research Training Group \textit{“GRK 2149: Strong and Weak 
	Interactions – from Hadrons to Dark Matter”}. We acknowledge the computer 
	resources provided by the \textit{Zentrum für Informationsverarbeitung} of the 
	University of Münster (PALMA II HPC cluster) and thank its staff for support. 
	We are grateful to our colleagues in the CLS initiative for producing the 
	gauge configuration ensembles used in this study.
	\vspace{-.3em}
	\input{kuberski_lat19.bbl}
\end{document}

%% file: kuberski_lat19.bbl
\providecommand{\href}[2]{#2}\begingroup\raggedright\endgroup